\documentclass[11pt,twoside]{article}
\usepackage[pdftex]{graphicx}
\usepackage{amsmath}
\usepackage{amssymb}
\usepackage{cite}

\begin{document}
\newcommand{\pst}{\hspace*{1.5em}}

\newcommand{\rigmark}{\em Journal of Russian Laser Research}
\newcommand{\lemark}{\em Volume 30, Number 5, 2009}

\newcommand{\be}{\begin{equation}}
\newcommand{\ee}{\end{equation}}
\newcommand{\bm}{\boldmath}
\newcommand{\ds}{\displaystyle}
\newcommand{\bea}{\begin{eqnarray}}
\newcommand{\eea}{\end{eqnarray}}
\newcommand{\ba}{\begin{array}}
\newcommand{\ea}{\end{array}}
\newcommand{\arcsinh}{\mathop{\rm arcsinh}\nolimits}
\newcommand{\arctanh}{\mathop{\rm arctanh}\nolimits}
\newcommand{\bc}{\begin{center}}
\newcommand{\ec}{\end{center}}

\thispagestyle{plain}

\label{sh}


\begin{center} {\Large \bf
\begin{tabular}{c}
IS THERE A PROBLEM WITH OUR HAMILTONIANS FOR
\\[-1mm]
QUANTUM NONLINEAR OPTICAL PROCESSES?
\end{tabular}
 } \end{center}

\bigskip

\bigskip

\begin{center} {\bf
Stephen M. Barnett}\end{center}

\medskip

\begin{center}
{\it
School of Physics and Astronomy, University of Glasgow, Glasgow G12 8QQ, U. K.}
\smallskip

Corresponding author e-mail:~~~stephen.barnett@glasgow.ac.uk\\
\end{center}

\begin{abstract}\noindent
The models we use, habitually, to describe quantum nonlinear optical processes have been remarkably successful
yet, with few exceptions, they each contain a mathematical flaw.  We present this flaw, show how it can be fixed and, in the 
process, suggest why we can continue to use our favoured Hamiltonians.  
\end{abstract}

\medskip

\noindent{\bf Keywords:}
quantum optics, nonlinear optics, model Hamiltonians, three-wave mixing.


\section{Introduction: a dilemma}
\pst
Quantum nonlinear optics is now very well developed, with photonic devices such as parametric oscillators 
and spontaneous parametric downconverters playing the role of work-horses in experimental demonstrations of exotic
quantum phenomena (including entanglement) and in the advance of quantum information technology.  Of particular
relevance to the topic of this special issue are those associated with second and third order nonlinearities, corresponding
to the interaction between three or four fields.  We shall see that the natural and widely employed quantum descriptions
of these processes rely on Hamiltonians with spectra that are unbounded from below \cite{Stig}.  There are very good 
mathematical and, indeed physical, reasons for doubting the validity of such an unbounded Hamiltonian and these 
doubts, in turn, challenge our confidence in our understanding of quantum nonlinear optical processes.  The problem 
is an old one, although perhaps not well known, and it is for this reason, primarily, that this article cites mostly books rather 
than original papers; tracking down the full set of relevant papers published over the last 30 or so years would be an 
exhausting challenge.

Let us begin by presenting the problem as simply as possible.  To this end we consider a simple model of an
intracavity optical parametric oscillator in which a nonlinear crystal mediates the reversible transformation of 
single photons from mode $a$ into a pair of photons, one each in modes $b$ and $c$.  A simple Hamiltonian used to
describe this process might be of the form
\begin{equation}
\label{Eq1}
\hat{H} = \omega_a \hat{a}^\dagger \hat{a} + \omega_b \hat{b}^\dagger \hat{b} + \omega_c \hat{c}^\dagger \hat{c}
+ \kappa\left(\hat{a}^\dagger \hat{b} \hat{c} + \hat{b}^\dagger\hat{c}^\dagger a\right) ,
\end{equation}
where $\omega_a = \omega_b + \omega_c$ and, as is common practice in quantum optics, we have chosen units 
such that $\hbar = 1$.  There is a very large variety of models of this form, these include four-wave 
processes in which there is an additional mode, with annihilation operator $\hat{d}$, and for which $\hat{a}$ is replaced 
by $\hat{a}\hat{d}$ in the interaction term, models
in which one or more modes are strong and treated classically and also models using continuum modes.  These are often
supplemented by driving terms and the ubiquitous losses and noise.  The treatment of these is now the domain of specialist
textbooks devoted to quantum optics \cite{Louisell,Perina,Meystre,Walls,LMandel,Scully,Methods,Bachor,Loudon,Carmichael}.
That our simple Hamiltonian has a spectrum that is unbounded from below is most readily demonstrated by evaluating the
expectation value of $\hat{H}$ for the three-mode coherent state $|\alpha\rangle_a|\beta\rangle_b|\gamma\rangle_c$ \cite{Stig}
\begin{equation}
\label{Eq2}
\langle\hat{H}\rangle = \omega_a |\alpha|^2 + \omega_b |\beta|^2 + \omega_c |\gamma|^2
+ \kappa\left(\alpha^*\beta\gamma + \beta^*\gamma^*\alpha\right)
\end{equation}
where we have used the familiar eigenstate property of the coherent states: $\hat{a}|\alpha\rangle = \alpha|\alpha\rangle$
\cite{Methods}.  This expectation value can take any value and, in particular, any negative value as may readily be seen by 
setting $|\alpha| = |\beta| = |\gamma|$ and choosing the phases of these such that the interaction term is negative:
\begin{equation}
\label{Eq3}
\langle\hat{H}\rangle = (\omega_a  + \omega_b + \omega_c)|\alpha|^2 - 2|\kappa ||\alpha|^3 ,
\end{equation}
which tends to $-\infty$ for large $|\alpha|$.  The expectation value of the Hamiltonian cannot be less than its lowest energy
eigenvalue and it follows, therefore, that the Hamiltonian has a spectrum that is unbounded from below.  

There are very good reasons for being suspicious of and even rejecting Hamiltonians with no ground state.  Perhaps the most telling
of these is the possibility, were such a system to be realised, of extracting unbounded amounts of energy associated with the
decay of the system to every lower energy states.


\section{Resolution I: Restricted state space}
\pst
The first thing to notice about the above argument is that the runaway behaviour towards negative energy eigenstates 
sets in at very high photon numbers corresponding to high optical electric field strengths.  As an indication of this
we can write our polarisation as a nonlinear function of the electric field \cite{Bob,Geoff}:
\begin{equation}
\label{Eq4}
P = \varepsilon_0\left( \chi^{(1)}E + \chi^{(2)}E^2 + \cdots \right)
\end{equation}
Typical values of the nonlinear susceptibility, $\chi^{(2)}$, are in the range $10^{-11}$ to $10^{-12}{\rm mV}^{-1}$ \cite{Bob} 
and this suggests that we need an optical electric field strength in the region of perhaps $10^{12}{\rm Vm}^{-1}$, corresponding
to an intensity of the order of $10^{18}{\rm Wcm}^{-2}$ before the nonlinear susceptibility dominates and perhaps leads to 
the problems indicated.  So one might very well take the view that the problem does not arise in the experimental
regime of interest.  This is not quite satisfactory, however, unless we can show that the dynamical evolution of the modes
cannot take us into the regime in which the unbounded negative-energy eigenspace occurs.

There is, thankfully, a natural set of conservation relations that apply and these ensure that if our initial state has no overlap
with the troublesome negative energy eigenstates then it will not acquire one.  It may be shown, either by inspection of the 
Hamiltonian or by explicit calculation, that there are three conserved quantities corresponding to the operators
\begin{eqnarray}
\label{Eq5}
\hat{M}_1 &=& \hat{b}^\dagger\hat{b} + \hat{a}^\dagger\hat{a}  \nonumber \\
\hat{M}_2 &=& \hat{c}^\dagger\hat{c} + \hat{a}^\dagger\hat{a} ,
\end{eqnarray}
with the third following from these:
\begin{equation}
\label{Eq6}
\hat{M}_3 = \hat{c}^\dagger\hat{c} - \hat{b}^\dagger\hat{b}  .
\end{equation}
These three are the operator analogues of the Manley-Rowe relations familiar from classical nonlinear optics \cite{Bob,Geoff}.
It follows that we can divide the state-space into non-interacting blocks, each characterised by a pair of positive integers,
$M_1$ and $M_2$ corresponding to the first of our two conserved quantities.  It then follows that within each block,
the number of photons in any mode can never exceed the largest of these two integers.  This procedure provides a
very natural way to solve for the dynamics of the three-mode state and of its mathematically equivalent model of
a number of two level-atoms interacting cooperatively with a single cavity mode \cite{Bonifacio,Barakat,Knight}.

In light of the above observations it is interesting, at least mathematically, to ask how it is that a Hamiltonian with 
photon-number conservation laws can lead to an energy eigenvalue spectrum that is unbounded from below, as surely
adding more photons will increase the energy.  In fact this is not so as may readily be seen by the following estimate.
Let us suppose that we have a low-energy eigenstate in which all three modes have roughly $N$ photons but with some
small variations in the superposition of photon-number product states.  For this state we can estimate the energy 
eigenvalue by replacing each of the creation and annihilation operators by $\sqrt{N}$ and suitably selecting the 
phases in the superposition such that the contribution from the interaction term is negative.  We find
\begin{equation}
\label{Eq7}
E_N \approx 2N\omega_a - 2|\kappa|N^{3/2} ,
\end{equation}
so that the energy becomes negative for $\sqrt{N} > \omega_a/|\kappa|$.  It will become increasingly negative as we 
\emph{increase} $N$ so that adding photons in this regime will \emph{reduce} the overall energy for this state.  If we restrict
our state-space to contain only photon numbers very much less than this value then using our coupled-mode
Hamiltonian will not get us into the difficulties associated with much higher photon numbers.  This means that we will
avoid problems if we restrict our Hamiltonian as being valid \emph{only} for photon numbers below some upper limit
and restrict our state space to be spanned only by photon numbers less than this upper value.

\begin{figure}[htbp] 
\centering
\includegraphics[width=10cm]{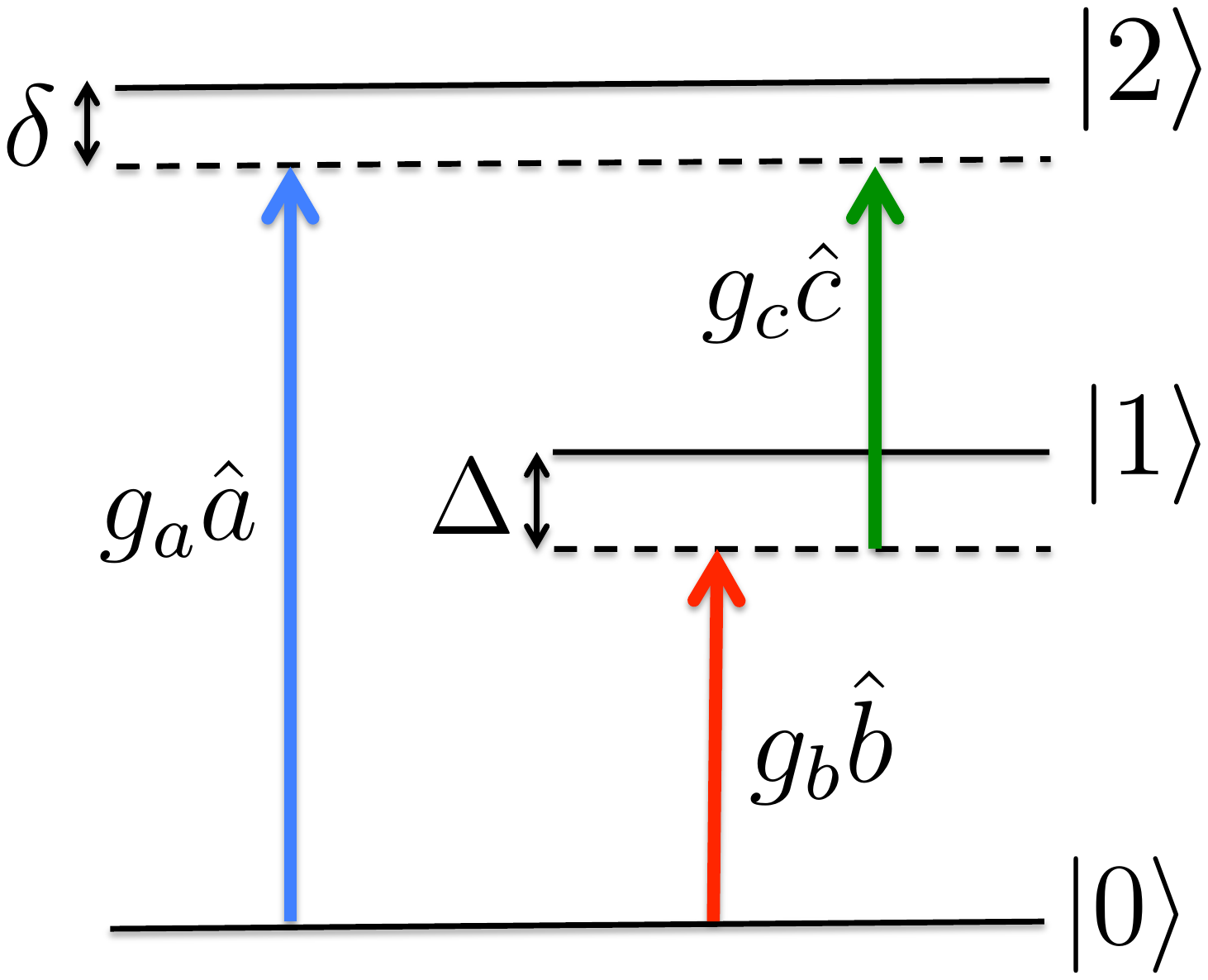}
\caption{Electronic energy level scheme and couplings for a microscopic theory of our nonlinear optical process.
Here the detunings $\delta$ and $\Delta$ are $\delta = \mathcal{E}_2 - \mathcal{E}_0 - \omega_a$
and $\Delta = \mathcal{E}_1 - \mathcal{E}_0 - \omega_b$.} 
\label{fig:figure1}
\end{figure}


\section{Resolution II: Higher-order processes}
\pst
The resolution presented in the preceding section is somewhat mathematical in nature and, as a counterbalance, we 
present here a more physical line of reasoning.  We have seen that the problem of unbounded negative energy 
eigenvalues arises at high field strengths, corresponding to very high photon numbers and this suggests that 
higher-order nonlinear optical processes will come into play before the problem is reached.  If so, then the 
physical resolution will be a more accurate Hamiltonian that does not have the problem of unbounded negative-energy
eigenstates. This does indeed turn out to be the case.

Let us consider a three-level atomic model with a ground state $|0\rangle$ and two excited states, $|1\rangle$ and 
$|2\rangle$, coupled by our three optical modes, as depicted in Fig. \ref{fig:figure1}\footnote{Strictly speaking we 
need an additional electric field to mix the parities of the energy levels so that the pattern of transitions are all allowed.}.
Such level schemes form the basis of microscopic calculations of nonlinear susceptibilities and can be found in many
texts on nonlinear optics \cite{Bob,Geoff,Yariv,Schubert,Butcher,Newell,PMandel}.
It suffices, for our purposes, to consider just a single atom as our nonlinear medium; including many atoms to form
a nonlinear medium presents no special difficulties but would add an unnecessary complication.  Our Hamiltonian
has the form
\begin{equation}
\label{Eq8}
\hat{H} = \hat{H}_{\rm F} + \hat{H}_{\rm A} + \hat{V} .
\end{equation}
where the component parts are
\begin{eqnarray}
\label{Eq9}
\hat{H}_{\rm F} &=& \omega_a \hat{a}^\dagger \hat{a} + \omega_b \hat{b}^\dagger \hat{b} + 
\omega_c \hat{c}^\dagger \hat{c} \nonumber \\
\hat{H}_{\rm A} &=& \sum_{i=0}^2 \mathcal{E}_i |i\rangle\langle i|   \nonumber \\
\hat{V} &=& g_a\left(|2\rangle\langle 0|\hat{a} + \hat{a}^\dagger|0\rangle\langle 2|\right)
+ g_c\left(|2\rangle\langle 1|\hat{c} + \hat{c}^\dagger|1\rangle\langle 2|\right)  \nonumber \\
& & \qquad \qquad + g_b\left(|1\rangle\langle 0|\hat{b} + \hat{b}^\dagger|0\rangle\langle 1|\right) .
\end{eqnarray}
It is interesting to note that this Hamiltonian provides a physical picture of the origin of entanglement generated between
modes $b$ and $c$.  If mode $a$ starts in a coherent state then the interaction with the atom imprints a phase
from mode $a$ onto the probability amplitude for energy level $2$ so that there is a coherence induced between levels
$0$ and $2$.  This coherence is transferred to modes $b$ and $c$ on transition to the atomic ground state so as to produce
a non-vanishing expectation value $\langle\hat{b}\hat{c}\rangle$ even though $\langle\hat{b}\rangle = 0 = \langle\hat{c}\rangle$,
which is a result of the entanglement between the modes.  We may view this as a manifestation of interference
between the two possible excitation pathways between the states $|0\rangle$ and $|2\rangle$
\cite{Krinitzky,Buckle,Bruce1,Swain,StigKalle,Bruce2}.

We can recover our initial nonlinear optical Hamiltonian by applying perturbation theory, up to third order in the 
atom-field coupling, to the atomic ground state.  If we then eliminate the ground state then we find 
\begin{equation}
\label{Eq10}
\hat{H} = \omega_a \hat{a}^\dagger \hat{a} + \omega_b \hat{b}^\dagger \hat{b} + 
\omega_c \hat{c}^\dagger \hat{c} -\frac{g_a^2}{\delta}\hat{a}^\dagger\hat{a} - \frac{g_b^2}{\Delta}\hat{b}^\dagger\hat{b}
+ \frac{g_ag_bg_c}{\delta\Delta}\left(\hat{a}^\dagger \hat{b} \hat{c} + \hat{b}^\dagger\hat{c}^\dagger a\right) .
\end{equation}
The fourth and fifth terms in this Hamiltonian correspond to Stark shifts of the ground state, but appear here as a
modification of frequencies of the cavity modes.  These account for the effective refractive index associated with the
presence of the atom and we can include these in the frequencies for the modes to give
\begin{equation}
\label{Eq11}
\hat{H} = \omega_a \hat{a}^\dagger \hat{a} + \omega_b \hat{b}^\dagger \hat{b} + 
\omega_c \hat{c}^\dagger \hat{c} 
+ \frac{g_ag_bg_c}{\delta\Delta}\left(\hat{a}^\dagger \hat{b} \hat{c} + \hat{b}^\dagger\hat{c}^\dagger a\right) ,
\end{equation}
which becomes our initial Hamiltonian, equation (\ref{Eq1}), if we set $\kappa = g_ag_bg_c/\delta\Delta$.  

Our perturbative approximation has reproduced the Hamiltonian which has the energy spectrum with 
no lower bound.  To show that this is a consequence of the approximations used to derive it we need
only show that the microscopic Hamiltonian from which it was derived, equation (\ref{Eq8}), does not 
have this property.  To this end we evaluate the expectation value of our microscopic Hamiltonian for
a general atomic state, $\sum_i c_i|i\rangle$, and the field coherent state 
$|\alpha\rangle_a|\beta\rangle_b|\gamma\rangle_c$.  For this state we find
\begin{eqnarray}
\label{Eq12}
\langle\hat{H}\rangle &=& \omega_a |\alpha|^2 + \omega_b |\beta|^2 + \omega_c |\gamma|^2
+ \sum_{i=0}^2 \mathcal{E}_i|c_i|^2 + g_a(\alpha c_0c^*_2 + \alpha^*c_2c^*_0) \nonumber \\
& & \qquad \qquad + g_c(\alpha c_1c^*_2 + \alpha^*c_2c^*_1) + g_b(\alpha c_0c^*_1 + \alpha^*c_1c^*_0) .
\end{eqnarray}
This expectation value tends to $+\infty$ as the amplitudes of the coherent states become large and
the problem of unbounded negative eigenvalues does not arise.  We conclude that this unphysical 
behaviour arises as a consequence of extending a perturbative theory beyond the bounds of its validity.
It is not difficult to see where this breakdown occurs; if the couplings between the atomic levels become
large compared with the detunings (for example $g_a\sqrt{N_a} \gg \delta$) then we start to find a significant
probability for the atom to be found in one its excited states and this invalidates the atomic ground-state
assumption used in deriving the approximate Hamiltonian, equation (\ref{Eq11}).


\section{Conclusion}
\pst
The problem we have identified with the simple Hamiltonian, equation (\ref{Eq1}) applies to a very wide range
of such model Hamiltonians used to describe quantum effects in nonlinear optics, including those used with
great success to describe the generation of entangled states.  That we can continue to use these model Hamiltonians
with confidence comes from the fact that the dynamics predicted by these models cannot enter the regime in which
unphysical behaviour would emerge.  That this is true mathematically is ensured by the conservation of the quantities 
$\hat{M}_1$ and $\hat{M}_2$ given in equation (\ref{Eq5}), which mean, in turn, that if we ensure that the maximum 
photon number in our analysis is sufficiently small then we can be confident in the predictions made using our model
Hamiltonian.  If we do push the model towards higher photon numbers, then we will, at some stage, need to abandon 
our simple multi-mode interaction term and include, explicitly, the dynamics of the nonlinear medium.

Perhaps we should, if only occasionally, acknowledge the fact that the Hamiltonians we use habitually in modelling
nonlinear optical processes should be used with caution \cite{Stig}, and that they form a valid description only when 
operating in a restricted state space in which the photon number is not too large.\

\section*{End note}

Stig Stenholm was a wonderful man, a brilliant and ingenious physicist, a scholar, a caring nurturer of young scientists and, perhaps
above all, an inquisitive, far-seeing and deep thinker.  We worked together for many years, albeit intermittently, and I owe to him
far more than these few words can adequately express.  Looking back, I find it surprising that we published only nine papers together
 \cite{Stig1,Stig2,Stig3,Stig4,Stig5,Stig6,Stig7,Stig8,Stig9}, but these papers constitute only a very small part of the fields we explored, 
 in physics, mathematics and, latterly, in philosophy, especially the philosophical foundations of quantum theory \cite{Bohmnote}.

Let me conclude by quoting the final paragraph from Stig's last book, {\it The Quest for Reality} \cite{Quest}, in which he sought to
reconcile distinct philosophical views of quantum theory.  He wrote:

{\it The situation is far from satisfactory, but it may, in the end, be the best world image available to our limited human intellect.
If that is so, we have to be grateful for what we get.  Chasing rainbows has never uncovered the treasures.  But the display of 
colors is magnificent.}

This sums up the man I knew rather well: ``you never really finish a problem", he once told me, but the fun is in the challenge to
understand.


\section*{Acknowledgments}
\pst
Let me put on record the debt of gratitude that I owe to Stig Stenholm, as mentor, inspiration, poser of challenging puzzles
(like the one addressed here) and, most importantly, as a friend.  This work was supported by a Royal Society Research Professorship
(RP150122).



\begin{thebibliography}{99}

\bibitem{Stig}
S. Stenholm  circa (1986) private communication.  His statement, if I remember correctly, was: ``Does it bother you that your Hamiltonian 
has a spectrum that is unbounded from below?".  The very same question was posed soon afterwards, also, by K. Rz\c{a}\.{z}ewski
after the presentation of a conference paper.

\bibitem{Louisell}
W. H. Louisell, {\it Radiation and Noise in Quantum Electronics}, McGraw-Hill, New York (1964).

\bibitem{Perina}
J. Pe\v{r}ina, {\it Quantum Statistics of Linear and Nonlinear Optical Phenomena}, Kluwer, Dordrecht (1991).

\bibitem{Meystre}
P. Meystre and D. F. Walls (eds.), {\it Nonclassical Effects in Quantum Optics}, American Institute of Physics, New York (1991).
This is a reprint collection a important early papers in quantum optics, many of them dealing with quantum nonlinear optics.

\bibitem{Walls}
D. F. Walls and G. J. Milburn, {\it Quantum Optics}, Springer-Verlag, Berlin (1994).

\bibitem{LMandel}
L. Mandel and E. Wolf, {\it Optical Coherence and Quantum Optics}, Cambridge University Press, Cambridge(1995).

\bibitem{Scully}
M. O. Scully and M. S. Zubairy, 1997 {\it Quantum Optics}, Cambridge University Press, Cambridge (1997).

\bibitem{Methods}
S. M. Barnett and P. M. Radmore, 1997 \emph{Methods in Theoretical Quantum Optics}, Oxford University Press, Oxford (1997).

\bibitem{Bachor}
H.-A. Bachor,1998 {\it A Guide to Experiments in Quantum Optics}, Wiley-VCH, Weinheim (1998).

\bibitem{Loudon}
R. Loudon, {\it The Quantum Theory of Light}, 3rd ed. Oxford University Press, Oxford (2000).

\bibitem{Carmichael}
H. J. Carmichael, {\it Statistical Methods in Quantum Optics 2}, Springer, Berlin (2008).

\bibitem{Bob}
R. W. Boyd, {\it Nonlinear Optics}, 3rd ed. Elsevier, Amsterdam (2008).

\bibitem{Geoff}
G. New, {\it Introduction to Nonlinear Optics}, Cambridge University Press, Cambridge (2011).

\bibitem{Bonifacio}
R. Bonifacio and G. Preparata, {\sl Lett. Nuovo Cim.}, {\bf 1} 887 (1969).

\bibitem{Barakat}
D. F. Walls and R. Barakat, {\sl Phys. Rev. A}, {\bf 1} 446 (1970).

\bibitem{Knight}
S. M. Barnett and P. L. Knight P L {\sl Opt. Acta}, {\bf 31} 435 (1984);  Erratum: 1984 {\bf 31} 1203 (1984).

\bibitem{Yariv}
A. Yariv. {\it Quantum Electronics}, 2nd ed. Wiley, New York (1975).

\bibitem{Schubert}
M. Schubert and B. Wilhelmi, {\it Nonlinear Optics and Quantum Electronics}, Wiley, New York (1986).

\bibitem{Butcher}
P. N. Butcher and D. Cotter, {\it The Elements of Nonlinear Optics}, Cambridge University Press, Cambridge (1990).

\bibitem{Newell}
A. G. Newell and J. V. Moloney, {\it Nonlinear Optics}, Addison-Wesley, Redwood City CA (1992).

\bibitem{PMandel}
P. Mandel P, {\it Nonlinear Optics}, Wiley-VCH, Weinheim (2010).

\bibitem{Krinitzky}
S. P. Krinitzky and D. T. Pegg, {\sl Phys. Rev. A}, {\bf 33} 403 (1986).

\bibitem{Buckle}
S. J. Buckle, S. M. Barnett, P. L. Knight, M. A. Lauder and D. T. Pegg, {\it J. Mod. Opt.}, {\bf 33} 1129 (1986).

\bibitem{Bruce1}
B. W. Shore, {\it The Theory of Coherent Atomic Excitation: Volume 2 Multilevel Atoms and Incoherence},
Wiley, New York (1990).

\bibitem{Swain}
Z. Ficek and S. Swain, {\it Quantum Interference and Coherence}, Springer, New York (2004).

\bibitem{StigKalle}
S. Stenholm and K.-A. Suominen, {\it Quantum Approach to Informatics}, Wiley, Hoboken NY (2005).

\bibitem{Bruce2}
B. W. Shore 2011 {\it Manipulating Quantum Structures Using Laser Pulses}, Cambridge University Press, Cambridge (2011).

\bibitem{Stig1}
M.-A. Dupertuis, S. M. Barnett and S. Stenholm, {\sl J. Opt. Soc. Am. B.}, \textbf{4}, 1102 (1987).

\bibitem{Stig2}
M.-A. Dupertuis, S. M. Barnett and S. Stenholm, {\sl J. Opt. Soc. Am. B.}, \textbf{4}, 1124 (1987).

\bibitem{Stig3}
C. R. Gilson, S. M. Barnett and S. Stenholm, {\sl J. Mod. Opt.}, \textbf{34}, 949 (1987).

\bibitem{Stig4}
S. M. Barnett, S. Stenholm and D. T. Pegg, {\sl Opt. Commun.}, \textbf{73}, 314 (1989).

\bibitem{Stig5}
S. M. Barnett and S. Stenholm, {\sl J. Mod. Opt.}, \textbf{47}, 2869 (2000).

\bibitem{Stig6}
S. Franke-Arnold, E. Andersson, S. M. Barnett and S. Stenholm, {\sl Phys. Rev. A}, \textbf{63}, 052301 (2001).

\bibitem{Stig7}
S. M. Barnett and S. Stenholm, {\sl Phys. Rev. A}, \textbf{64}, 033808 (2001).

\bibitem{Stig8}
J. Salo, S. M. Barnett and S. Stenholm, {\sl Opt. Commun.}, \textbf{259}, 772 (2006).

\bibitem{Stig9}
S. Croke, S. M. Barnett and S. Stenholm, {\sl Ann. Phys (N. Y.)}, \textbf{323}, 893 (2008).

\bibitem{Bohmnote}
S. M. Barnett and S. Stenholm, {\sl Unpublished note}, Causal interpretation of quantum mechanics:
an alternative to Bohmian mechanics (2009).

\bibitem{Quest}
S. Stenholm, {\it The Quest for Reality}, Oxford University Press, Oxford (2011).


\end{thebibliography}
\end{document}